\documentclass[aps,floatfix,twocolumn,showpacs,showkeys,preprintnumbers,amsmath]{revtex4}

\usepackage{dcolumn}
\usepackage{bm}
\usepackage{graphicx}
\graphicspath{{Images/}}

\begin{document}

\title{Nuclear symmetry energy and the role of the tensor force}
\author{Isaac Vida\~na$^1$, Artur Polls$^2$ and Constan\c{c}a Provid\^{e}ncia$^1$}
\affiliation{$^1$Centro de F\'{i}sica Computacional, Department of Physics, University of Coimbra, PT-3004-516
Coimbra, Portugal}
\affiliation{$^2$Departament d'Estructura i Constituents de la Mat\`eria and
Institut de Ci\`encies del Cosmos,
Universitat de Barcelona, Avda. Diagonal 647, E-08028 Barcelona, Spain}

\newcommand{\m}{\multicolumn}
\renewcommand{\arraystretch}{1.2}

\begin{abstract}

Using the Hellmann--Feynman theorem we analyze the contribution of the different terms of the nucleon-nucleon
interaction to the nuclear symmetry energy $E_{sym}$ and the slope parameter $L$. The analysis is performed
within the microscopic Brueckner--Hartree--Fock approach using the Argonne V18 potential plus the 
Urbana IX three-body force. We find that the main contribution to $E_{sym}$ and $L$
is due to the tensor component of the nuclear force.

\end{abstract}


\maketitle  


The nuclear symmetry energy, defined as the difference between the energies of neutron and symmetric matter, 
and in particular its density dependence, is a crucial ingredient to understand many important properties
of isospin-rich nuclei and neutron stars \cite{varan05,li08,steiner05}.  Experimental information on 
the density dependence of the symmetry energy $E_{sym}(\rho)$ below, close to and above saturation density
$\rho_0$ can be obtained from the analysis of data of isospin difusion measurements \cite{chen05}, 
giant \cite{garg07} and pygmy resonances \cite{pygmy}, isobaric analog states \cite{da09}, isoscaling 
\cite{shetty07} or meson production in heavy ion collisions \cite{li05b, fuchs06}. Accurate measurements of the neutron
skin thickness $\delta R=\sqrt{\langle R_n^2 \rangle}-\sqrt{\langle R_p^2\rangle}$ in heavy nuclei,
via parity-violating electron scattering experiments \cite{prex,roca11} or by means of antiprotonic atom
data \cite{brown07,centelles09}, can also help to constraint $E_{sym}(\rho)$, since its derivative its 
strongly correlated with $\delta R$ \cite{brown}. Additional information on $E_{sym}(\rho)$ can be extracted
from the astrophysical observations of compact
objects which open a window into both the bulk and microscopic properties of nuclear matter at extreme
isospin asymmetries \cite{steiner05}. In particular, the characterization of the core-crust
transition in neutron stars \cite{horo01,xu09,vi09,mou10,du10}, or the analysis of power-law correlations, such as the relation
between the radius of a neutron star and the equation of state \cite{lattimer01} can put stringent constraints
on $E_{sym}(\rho)$. Theoretically $E_{sym}(\rho)$ has been determined using both
phenomenological and microscopic many-body approaches. Phenomenological approaches, either relativistic
or non-relativistic, are based on effective interactions that are frequently built to reproduce
the properties of nuclei \cite{stone07}. Since many of such interactions are built to describe systems close to the
symmetric case, predictions at high asymmetries should be taken with care.  Skyrme--Hartree--Fock \cite{shf} and
relativistic mean field \cite{rmt} calculations are the most popular ones among them. Microscopic approaches start
from realistic nucleon-nucleon (NN) interactions that reproduce the scattering and bound state properties
of the free two-nucleon system and include naturally the isospin dependence \cite{muether99a}. The 
in-medium correlations are then built using many-body techniques that microscopically
account for isospin asymmetric effects such as, for instance, the difference in the Pauli blocking factors
of neutrons and protons in asymmetric matter. Among this type of approaches the most popular ones are
the Brueckner--Bethe--Goldstone (BBG) \cite{bbg} and the Dirac--Brueckner--Hartree--Fock (DBHF) \cite{dbhf} theories, 
the variational method \cite{var}, the correlated basis function (CBF) formalism \cite{cbf}, the self-consistent Green's function
technique (SCGF) \cite{scgf} or, recently, the V$_{low k}$ approach \cite{vlowk}. Nevertheless, in spite of the experimental \cite{lynch11} and 
theoretical \cite{li11} efforts carried out to study the properties of isospin-asymmetric nuclear systems, $E_{sym}(\rho)$
is still uncertain. Its value $E_{sym}$ at saturation is more or less well established ($\sim 30$ MeV), and its behavior
below saturation is now much better known \cite{tsang11}. However, for densities above $\rho_0$, $E_{sym}(\rho)$ is not 
well constrained yet, and the predictions from different approaches strongly diverge. Why $E_{sym}(\rho)$
is so uncertain is still an open question whose answer is related to our limited knowledge of the
nuclear force, and in particular of its spin and isospin dependence 
\cite{panda72,wiringa88,brown90,bombaci91,dieperink03,xu10,xu11,sammarruca11}. 


In this letter we analyze the contribution of the different terms of the NN interaction
to $E_{sym}$ and the slope parameter $L=3\,\rho_0\,(\partial E_{sym}(\rho)/\partial \rho)_{\rho_0}$.
The analysis is carried out with the help of the Hellmann--Feynman theorem \cite{hellmann} within the framework
of the microscopic Brueckner--Hartree--Fock (BHF) approach \cite{bbg}. We employ the Argonne V18 (Av18) potential
\cite{wiringa95} supplemented with the Urbana IX three-body force \cite{pudliner95} which for the use
in the BHF approach is reduced to an effective two-body density-dependent force by averaging
over the third nucleon \cite{loiseau71}. We find  that the tensor term of the nuclear force
gives the largest contribution to both $E_{sym}$ and $L$.


The BHF approch is the lowest order of the BBG many-body theory \cite{bbg}. 
In this theory, the ground state energy of nuclear matter is evaluated in terms of the
so-called {\it hole-line expansion}, where the perturbative diagrams are grouped according to the number of independent
hole-lines. The expansion is derived by means of the in-medium two-body scattering $G$-matrix. The $G$-matrix, that takes 
into account the effect of the Pauli principle on the scattered particles and the in-medium potential felt by each nucleon,
has a regular behavior even for strong short-range repulsions, and it describes the effective interaction between two 
nucleons in the presence of a surrounding medium. In the BHF approach, the energy is given by the sum of only 
{\it two-hole-line} diagrams including the effect of two-body correlations through the $G$-matrix. It has been shown 
by Song {\it et al.,} \cite{song98} that the contribution to the energy from {\it three-hole-line} diagrams (that account
for the effect of three-body correlations) is minimized when the so-called continous prescription \cite{jeukenne76} is adopted 
for the in-medium potential, which is a strong indication of the convergence of the hole-line expansion. We adopt this
prescription in our calculation.

\begin{center}
\begin{table}[t!]
\begin{tabular}{lrrrr}
\hline
\hline
 & $E_{NM}$ & $E_{SM}$ & $E_{sym}$ & $L$ \\
\hline
 $\langle T \rangle$                       &  $53.321$ &  $54.294$  &  $-0.973$  & $14.896$\\
 $\langle V \rangle$                       &  $-34.251$ &  $-69.524$ &  $35.273$  & $51.604$\\
 Total                                     &  $19.070$ &  $-15.230$ &  $34.300$  & $66.500$ \\
\hline
\hline
\end{tabular}
\caption{Kinetic $\langle T \rangle$ and potential $\langle V \rangle$ contributions
to $E_{NM}$, $E_{SM}$, $E_{sym}$ and $L$. Units are given in MeV.}
\label{tab1}
\end{table}
\end{center}

The BHF approach does not give access to the separate contributions of the kinetic and potential energy in the 
correlated many-body state, because it does not provide the correlated many-body wave function $|\Psi\rangle$. However, it 
has been recently shown \cite{muether99} that the Hellmann--Feyman theorem \cite{hellmann} can be used to calculate the 
ground state expectation values of both contributions from the derivative of the total energy with respect to a
properly introduced parameter. Writing the nuclear matter Hamiltonian as $H=T+V$, and defining a $\lambda$ dependent 
Hamiltonian $H(\lambda)=T+\lambda V$, the expectation value of the potential energy is given as 
\begin{equation}
\langle V \rangle \equiv \frac{\langle \Psi |V| \Psi \rangle}{\langle \Psi|\Psi \rangle}=\left(\frac{dE}{d\lambda}\right)_{\lambda=1} \ .
\label{eq:hft}
\end{equation}
Then, the kinetic energy contribution $\langle T \rangle$ can be obtained simply by subtracting $\langle V \rangle$
from the total energy E.


In Table \ref{tab1} we show the kinetic and potential contributions to the energy of neutron matter $E_{NM}$, symmetric matter $E_{SM}$, 
$E_{sym}$ and L at saturation ($\rho_0=0.187$ fm$^{-3}$ in our calculation). We note that the kinetic contribution to $E_{sym}$ is very
small and negative. This is in contrast with the result for a free Fermi gas (FFG), whose contribution at $\rho_0$ is $\sim 14.4$ MeV. A similar result has 
been recently found by Xu and Li \cite{xu11}. According to these authors, this is due to the strong isospin-dependence of 
the short-range NN correlations (SRC) induced by the tensor force. They have shown, in fact, that the increase  of the kinetic energy of 
symmetric matter due to SRC is much larger than that of neutron matter, the kinetic part of the symmetry energy becoming then
negative. We also note that the kinetic contribution to $L$ is smaller than the corresponding one of the FFG ($L^{FFG}\sim 29.2$ MeV). 
The major contribution to both $E_{sym}$ and $L$ is due to the potential part.  Note that, in fact, this contribution is
practically equal to the total value of $E_{sym}$ and it represents $\sim 78\%$ of $L$. 

\begin{center}
\begin{table}[t!]
\begin{tabular}{lrrrr}
\hline
\hline
Partial wave & $E_{NM}$ & $E_{SM}$ & $E_{sym}$ & $L$ \\
\hline
 $^1S_0$   &  $-23.070$        & $-19.660$  & $-3.410$   & $-3.459$     \\
 $^3S_1$   &  $0$              & $-45.810$  & $45.810$   & $71.855$    \\ 

 $^1P_1$   &  $0$              & $4.904$    & $-4.904$   & $-18.601$    \\
 $^3P_0$   &  $-5.321$         & $-4.029$   & $-1.292$   & $-1.898$    \\
 $^3P_1$   &  $16.110$         & $10.720$   & $5.390$    & $21.949$    \\
 $^3P_2$   &  $-16.000$        & $-9.334$   & $-6.666$   & $-21.168$   \\

 $^1D_2$   &  $-5.956$         & $-3.201$   & $-2.755$   & $-11.033$    \\
 $^3D_1$   &  $0$              & $0.981$    & $-0.981$   & $-3.739$    \\
 $^3D_2$   &  $0$              & $-3.982$   & $3.982$    & $16.601$   \\
 $^3D_3$   &  $0$              & $-0.798$   & $0.798$    & $4.895$   \\

 $^1F_3$   &  $0$              & $0.694$    & $-0.694$   & $-3.348$    \\
 $^3F_2$   &  $-0.695$         & $-0.229$   & $-0.466$   & $-1.799$   \\
 $^3F_3$   &  $2.000$          & $0.821$    & $1.179$    & $4.883$   \\
 $^3F_4$   &  $-0.796$         & $-0.194$   & $-0.602$   & $-3.239$   \\

 $^1G_4$   &  $-0.812$         & $-0.247$   & $-0.565$   & $-3.036$    \\
 $^3G_3$   &  $0$              & $-0.001$   & $0.001$    & $0.441$   \\
 $^3G_4$   &  $0$              & $-0.213$   & $0.213$    & $0.449$   \\
 $^3G_5$   &  $0$              & $-0.057$   & $0.057$    & $0.650$   \\

 $^1H_5$   &  $0$              & $0.029$    & $-0.029$   & $0.107$    \\
 $^3H_4$   &  $0.033$          & $0.040$    & $-0.007$   & $0.232$   \\
 $^3H_5$   &  $0.225$          & $-0.033$   & $0.258$    & $0.968$   \\
 $^3H_6$   &  $0.043$          & $0.034$    & $0.009$    & $0.144$   \\

 $^1I_6$   &  $-0.082$         & $0.023$    & $-0.105$   & $-0.591$    \\
 $^3I_5$   &  $0$              & $-0.029$   & $0.029$    & $0.342$   \\
 $^3I_6$   &  $0$              & $0.067$    & $-0.067$   & $-0.819$   \\
 $^3I_7$   &  $0$              & $-0.021$   & $0.021$    & $0.239$   \\

 $^1J_7$   &  $0$              & $-0.027$   & $0.027$    & $0.385$    \\ 
 $^3J_6$   &  $0.044$          & $0.020$    & $0.024$    & $0.283$   \\
 $^3J_7$   &  $-0.062$         & $-0.060$   & $-0.002$   & $-0.313$   \\
 $^3J_8$   &  $0.036$          & $0.014$    & $0.022$    & $0.242$   \\

 $^1K_8$   &  $0.031$          & $0.021$    & $0.010$    & $0.169$    \\
 $^3K_7$   &  $0$              & $-0.011$   & $0.011$    & $0.138$   \\
 $^3K_8$   &  $0$              & $0.038$    & $-0.038$   & $-0.491$   \\

 $^3L_8$   &  $0.021$          & $0.006$    & $0.015$    & $0.166$   \\
\hline
\hline
\end{tabular}
\caption{Partial wave decomposition of the potential part of 
$E_{NM}$, $E_{SM}$, $E_{sym}$ and $L$. Units are given in MeV.}
\label{tab2}
\end{table}
\end{center}

\begin{center}
\begin{table}[h!]
\begin{tabular}{lrrrr}
\hline
\hline
$(S,T)$ & $E_{NM}$ & $E_{SM}$ & $E_{sym}$ & $L$ \\
\hline
 $(0,0)$    &  $0$        & $ 5.600$   & $-5.600$  & $-21.457$    \\
 $(0,1)$    &  $-29.889 $ & $-23.064$  & $-6.825$  & $-17.950$  \\
 $(1,0)$    &  $0$        & $-49.836$  & $49.836$  & $90.561$  \\
 $(1,1)$    &  $-4.362$   & $-2.224$   & $-2.138$  & $0.450$  \\
\hline
\hline
\end{tabular}
\caption{Spin (S) and isospin (T) channel decomposition of 
the potential part of $E_{NM}$, $E_{SM}$, $E_{sym}$ and $L$. Units are given in MeV.}
\label{tab3}
\end{table}
\end{center}


Tables \ref{tab2} and \ref{tab3} show the partial wave, and the spin (S) and isospin (T) channel decompositions of 
the potential part of $E_{NM}$, $E_{SM}$, $E_{sym}$ and $L$ at $\rho_0$. Contributions up to $J=8$ have been considered. We observe
that the spin-triplet ($S=1$) and isospin-singlet ($T=0$) channel, and in particular the $^3S_1$ wave, gives the largest contribution
to both $E_{sym}$ and $L$. This is due, as we explicitly show in the following, to the effect of the tensor component of nuclear force that dominates
the potential contribution to the symmetry energy and $L$, mainly through the $^3S_1-^3D_1$ channel. Note that this channel, which gives the
major contribution to the energy of symmetric matter, does not contribute to neutron matter. Note also that isospin-triplet ($T=1$)
channels give similar contributions  to both $E_{NM}$ and $E_{sym}$ which almost cancel out in $E_{sym}$. Similar arguments have
been pointed out by other authors \cite{panda72,wiringa88,brown90,bombaci91,dieperink03,xu10,xu11,sammarruca11}.


Next, we analyze the role played by the different terms of the nuclear force, particularly the one of the tensor, in the
determination of $E_{sym}$ and $L$. To such end we apply the Hellmann--Feynman theorem to the separate components of the 
Av18 potential and the Urbana IX three-body force. The Av18 potential has $18$ components of the form $v_p(r_{ij})O^p_{ij}$ with
\begin{eqnarray}
&&O^{p=1,18}_{ij}=1, \vec \tau_i \cdot \vec \tau_j, \vec \sigma_i \cdot \vec\sigma_j, 
(\vec \sigma_i \cdot \vec \sigma_j )(\vec \tau_i \cdot \vec \tau_j), \nonumber \\
&&S_{ij},S_{ij}(\vec \tau_i \cdot \vec \tau_j), \vec L \cdot \vec S,
\vec L \cdot \vec S(\vec \tau_i \cdot \vec \tau_j), L^2, \nonumber \\ 
&&L^2(\vec \tau_i \cdot \vec \tau_j), 
L^2(\vec \sigma_i \cdot \vec \sigma_j), L^2(\vec \sigma_i \cdot \vec \sigma_j)(\vec \tau_i \cdot \vec \tau_j), 
(\vec L \cdot \vec S)^2, \nonumber \\
&&(\vec L \cdot \vec S)^2(\vec \tau_i \cdot \vec \tau_j), T_{ij},(\vec \sigma_i \cdot \vec \sigma_j)T_{ij},
S_{ij}T_{ij},(\tau_{zi}+\tau_{z_j}) \nonumber
\label{eq:av18}
\end{eqnarray}
being $S_{ij}$ the usual tensor operator, $\vec L$ the relative orbital angular momentum, 
$\vec S$ the total spin of the nucleon pair, and $T_{ij}=3\tau_{zi}\tau_{zj}-\tau_i \cdot \tau_j$ the isotensor operator defined
analogously to $S_{ij}$. Note that the last four operators break the charge independence of the nuclear interaction.

\begin{center}
\begin{table}
\begin{tabular}{lrrrr}
\hline
\hline
 & $E_{NM}$ & $E_{SM}$ & $E_{sym}$ & $L$ \\
\hline
 $\langle V_1 \rangle $    &  $-31.212$ & $-32.710$  & $1.498$   & $-5.580$    \\
 $\langle V_{\vec\tau_i\cdot\vec\tau_j} \rangle$     &  $-4.957$  & $3.997$    & $-8.954$  & $-20.383$  \\
 $\langle V_{\vec\sigma_i\cdot\vec\sigma_j} \rangle$     &  $-0.319$  & $-0.382$   & $0.063$   & $2.392$    \\
 $\langle V_{(\vec\sigma_i\cdot\vec\sigma_j)(\vec\tau_i\cdot\vec\tau_j)} \rangle$     &  $-5.724$  & $-11.388$  & $5.664$   & $2.521$    \\
 $\langle V_{S_{ij}} \rangle$     &  $-0.792$  & $1.912$    & $-2.704$  & $-4.998$   \\
 $\langle V_{S_{ij}(\vec\tau_i\cdot\vec\tau_j)} \rangle$     &  $-4.989$  & $-37.592$  & $32.603$  & $47.095$   \\
 $\langle V_{\vec L\cdot \vec S} \rangle$     &  $-7.538$  & $-1.754$   & $-5.784$  & $-12.251$  \\
 $\langle V_{\vec L\cdot \vec S (\vec\tau_i\cdot\vec\tau_j)} \rangle$     &  $-2.671$  & $-6.539$   & $3.868$   & $3.969$    \\
 $\langle V_{L^2} \rangle$     &  $11.850$  & $13.610$   & $-1.760$  & $1.521$    \\
 $\langle V_{L^2(\vec\tau_i\cdot\vec\tau_j)} \rangle$  &  $-2.788$  & $0.270$    & $-3.058$  & $-14.262$  \\
 $\langle V_{L^2(\vec\sigma_i\cdot\vec\sigma_j)} \rangle$  &  $1.265$   & $1.383$    & $-0.118$  & $1.405$    \\
 $\langle V_{L^2(\vec\sigma_i\cdot\vec\sigma_j)(\vec\tau_i\cdot\vec\tau_j)} \rangle$  &  $0.051$   & $0.008$    & $0.043$   & $-0.341$   \\
 $\langle V_{(\vec L\cdot \vec S)^2} \rangle$  &  $4.194$   & $5.682$    & $-1.488$  & $-0.327$   \\
 $\langle V_{(\vec L\cdot \vec S)^2(\vec\tau_i\cdot\vec\tau_j)} \rangle$  &  $5.169$   & $-6.190$   & $11.359$  & $31.368$    \\
 $\langle V_{T_{ij}} \rangle$  &  $0.003$   & $0.039$    & $-0.036$  & $-0.022$   \\
 $\langle V_{(\vec\sigma_i\cdot\vec\sigma_j)T_{ij}} \rangle$  &  $-0.017$  & $-0.106$   &  $0.089$  & $0.042$    \\
 $\langle V_{S_{ij}T_{ij}} \rangle$  &  $0.004$   & $0.079$    &  $-0.075$ & $-0.124$   \\
 $\langle V_{(\tau_{z_i}+\tau_{z_j})} \rangle$  &  $-0.084$  & $-0.001$   &  $-0.083$ & $-0.331$   \\
\\
 $\langle U_1 \rangle$   &  $2.985$   &  $3.251$   &  $-0.266$ & $-0.630$  \\
 $\langle U_{(\vec\sigma_i\cdot\vec\sigma_j)(\vec\tau_i\cdot\vec\tau_j)} \rangle$  &  $2.254$   &  $3.999$   &  $-1.745$ & $-7.228$   \\
 $\langle U_{S_{ij}(\vec\tau_i\cdot\vec\tau_j)} \rangle$  &  $-0.935$  &  $-7.092$  &  $6.157$  & $27.768$   \\ 
\hline
\hline
\end{tabular}
\caption{Separate contributions to $E_{NM}$, $E_{SM}$, $E_{sym}$ and $L$ from the
various components of the Av18 potential (denoted as $\langle V_i \rangle$) and the
reduced Urbana force (denoted as $\langle U_i \rangle$). Units are given in MeV.}
\label{tab4}
\end{table}
\end{center}

As we said above, the Urbana IX three-body force is reduced to an effective density-dependent two-body force when used in the BHF 
approach. For simplicity, in the following we refer to it as reduced Urbana force. This force is made of $3$ components of the 
type $u_p(r_{ij},\rho)O^p_{ij}$ where $O^{p=1,3}_{ij}=1, (\vec\sigma_i\cdot\vec\sigma_j)(\vec\tau_i\cdot\vec\tau_j), S_{ij}(\vec\tau_i\cdot\vec\tau_j)$,
introducing additional central, $\sigma\tau$ and tensor terms 
(see {\it e.g.,} Baldo and Ferreira in Ref.\ \cite{loiseau71} for details).

The separate contributions to $E_{NM}$, $E_{SM}$, $E_{sym}$ and $L$ from the various components of the Av18 potential and 
the reduced Urbana force are given in Table \ref{tab4}. The contribution from the tensor component to $E_{sym}$ and
$L$ (contributions $\langle V_{S_{ij}} \rangle$ and $\langle V_{S_{ij}(\vec \tau_i \cdot \vec \tau_j)}\rangle$ from the
Av18 potential, and $\langle U_{S_{ij}(\vec \tau_i \cdot \vec \tau_j)}\rangle$ from the reduced Urbana force) is
$36.056$ MeV and $69.968$ MeV, respectively. These results clearly confirm that the tensor force gives the largest contribution
to both $E_{sym}$ and $L$. The contributions from the other components are either negligible, as for instance the contribution
from the charge symmetry breaking terms ($\langle V_{T_{ij}}\rangle, \langle V_{(\vec \sigma_i \cdot \vec \sigma_j)T_{ij}}\rangle, 
\langle V_{S_{ij}T_{ij}}\rangle$ and $\langle V_{(\tau_{zi}+\tau_{z_j})}\rangle $), or almost cancel out.

\vspace{0.25cm}
In summary, using the Hellmann--Feynman theorem we have evaluated the separate contribution of the different terms of the nuclear force
to the nuclear symmetry energy $E_{sym}$ and the slope parameter $L$. Our study has been done within the framework of the BHF
approach using the Av18 potential plus an effective density-dependent two-body force deduced from the Urbana IX three-body one.
Our results show that the potential part of the nuclear Hamiltonian gives the main contribution to both $E_{sym}$ and $L$. The 
kinetic contribution to $E_{sym}$ is very small and negative in agreement with the recent results of Xu and Li \cite{xu11}. 
We have performed a partial wave, and a spin-isospin channel decomposition of the potential part of $E_{sym}$ and $L$, showing that the major
contribution to them is given by the spin-triplet ($S=1$) and isospin-singlet ($T=0$) channel. This is due, as we have explicitly shown, to the dominant effect
of the tensor force which gives the largest contribution to both $E_{sym}$ and $L$. In conclusion, our results confirm the 
critical role of the tensor force in the determination of the symmetry energy and its density dependence. 

\vspace{0.25cm}
This work has been partially supported by FEDER and FCT (Portugal) under the projects PTDC/FIS/113292/2009, CERB/FP/109316/2009
and CERN/FP/116366/2010, the Consolider Ingenio 2010 Programme CPAN CSD2007-00042 and Grant No.  FIS2008-01661 from MEV and FEDER
(Spain) and Grant 2009GR-1289 from Generalitat de Catalunya (Spain), and by COMPSTAR, and ESF Research Networking Programme


\end{document}